\documentclass[reprint,superscriptaddress,amsmath,amssymb,aps,prb,longbibliography]{revtex4-2}
\usepackage[utf8]{inputenc}
\usepackage{graphicx}
\usepackage{dcolumn}% Align table columns on decimal point
\usepackage{bm}% bold math
\usepackage[colorlinks,citecolor=blue,linkcolor=blue,urlcolor=blue]{hyperref}
\usepackage{gensymb}
\begin{document}

% \title{Calculation of minimal energy path of anisotropic trimer and pyramaid structure in non-collinear Alexander-Anderson model using magnetic force theorem}
\title{Control of localized states of itinerant electrons and their magnetic interactions}
%\title{Magnetic dimer with itinerant electrons: \\
%plural solutions and non-collinear ground state}
\author{Yaxin Sun}
\affiliation{School of Science, Harbin Institute of Technology, Shenzhen, 518055, China}
\affiliation{Shenzhen Key Laboratory of Advanced Functional Carbon Materials Research and Comprehensive Application, Shenzhen 518055, China.}

\author{I. S. Lobanov}
%\thanks{These authors contributed equally.}
\affiliation{Faculty of Physics, ITMO University, 197101 St. Petersburg, Russia.}
 
\author{Jiahao Su}
\affiliation{School of Science, Harbin Institute of Technology, Shenzhen, 518055, China}
\affiliation{Shenzhen Key Laboratory of Advanced Functional Carbon Materials Research and Comprehensive Application, Shenzhen 518055, China.}

\author{Ho-Kin Tang}
\email{denghaojian@hit.edu.cn}
\affiliation{School of Science, Harbin Institute of Technology, Shenzhen, 518055, China}
\affiliation{Shenzhen Key Laboratory of Advanced Functional Carbon Materials Research and Comprehensive Application, Shenzhen 518055, China.}

\author{V. M. Uzdin}
\email{valery.uzdin@metalab.ifmo.ru}
\affiliation{Faculty of Physics, ITMO University, 197101 St. Petersburg, Russia.}
\affiliation{Faculty of Physics, St. Petersburg State University, 198504. St. Petersburg, Russia}

\date{\today}

\begin{abstract}

Controlling the magnetic properties of nanosystems by an electric field offers a number of advantages for spintronics applications. Using the noncollinear Alexander-Anderson model, we have shown that the interaction of localized magnetic moments formed by itinerant electrons strongly depends on the position of the d-level relative to the Fermi level, which determines the number of localized electrons.  Depending on this parameter, the ground state of the magnetic dimer can be ferromagnetic, antiferromagnetic, or noncollinear without the effects of spin-orbit interaction. The magnetic state can be controlled by shifting the d-level with an electric field, even without current flow. For a sufficiently large value of the hopping parameter between localized states there can be several self-consistent solutions with different values of magnetic moments. This opens up new possibilities for the manipulation of the magnetic structure of nanosystems. The results obtained lead to a new interpretation of the mechanisms of magnetization reversal, recording, and deleting of magnetic structures in tunneling spectroscopy experiments.

\end{abstract}
\pacs{Valid PACS appear here}

\maketitle
 
{\section{Introduction}
The magnetic properties of 3d transition metals, alloys, and embedded clusters are caused by itinerant electrons. In these systems, the magnetic moments of atoms in Bohr magnetons are not integers and can depend on local configuration, proximity to the surface, interface, or structural defects~\cite{coey2010magnetism}. Of particular interest is the formation of localized non-collinear magnetic structures, which can serve as bits of magnetic memory~\cite{Rimmler_NRM_2025}. Stability with respect to thermal fluctuations is a key consideration for these applications. However, the calculation of magnetic structures with atomic resolution within the framework of models featuring a continuous magnetization distribution, for example, by density functional theory (DFT), becomes computationally challenging even for systems with just a few tens of magnetic atoms. Consequently, Heisenberg-type models with localized magnetic moments are often used~\cite{Szilva_RMP_2023} to analyze the stability and magnetic properties of such $3d$ structures~\cite{Bessarab_Sci-rep_2018}. In this context, model parameters are typically chosen
to reproduce some characteristics obtained from density functional calculations~\cite{Szilva_RMP_2023, Bessarab_Sci-rep_2018, Hoffmann_PRB_2020}.
Based on this approach, in the present work we propose a mechanism by which variation in the d-electrons number $N$, associated with a shift of the d-level relative to the Fermi level, modifies the effective interaction between magnetic states, changes the corresponding parameters of discrete models, and allows control over magnetic ordering.

Typically, the parameters of the Heisenberg model can be determined by considering infinitesimal rotations of magnetic moments relative to equilibrium states in certain magnetic configurations. Such configurations can be a collinear ferromagnetic (FM), antiferromagnetic (AF) state, or a non-collinear localized magnetic structure. Generally speaking, the parameters obtained for the Heisenberg exchange depend on the configuration with respect to which small deviations of the moments are considered~\cite{Liechtenstein_JPF_1984,Szilva_RMP_2023}. Note that even for a magnetic dimer, for certain values of the parameters in the tight-binding model, the ground state is non-collinear~\cite{costa_PRL_2005}. In the correspondent generalized Heisenberg model, this indicates the need to introduce a biquadratic exchange interaction. A non-collinear structure arises even without taking into account the spin-orbit interaction, which in turn can lead to the appearance of the Dzyaloshinskii-Moriya interaction, which is responsible for formation of non-collinear chiral structures in magnetic systems~\cite{Szilva_RMP_2023,Dias_PRB_21,Cardias_PRB_22}.

Some previous studies claim that an adequate description of the magnetic structure and properties requires going beyond the Heisenberg approximation. This requires accounting for complex interactions such as multispin exchange and configuration-dependent effective interactions, which become significant in noncollinear magnetic regimes~\cite{Streib_PRB_22}.
However, the question of the ability of the localized electron model to reproduce all the features of the magnetism of itinerant electrons remains open. What peculiarities of the itinerant model are not accounted for in the generalized Heisenberg model but are important for describing the magnetic behavior and controlling magnetic interactions?

To study this issue, we will consider the simplest system of magnetic dimer within the itinerant non-collinear (NC) Alexander-Anderson (AA) model~\cite{Bessarab_PRB_2014} and investigate its properties. 

The article is organized as follows:
the second section considers a non-collinear magnetic dimer within the NC AA model and demonstrates that, in the mean-field approximation, self-consistent solutions in the canonical ensemble correspond to extrema of  energy as a function of the magnetic moment and the parameter determining the position of the d-level relative to the Fermi level. The case of a grand canonical ensemble is considered in the appendix. The third section presents self-consistent solutions for different numbers of d-electrons and discusses the stability of these solutions. Later, the fourth section considers the non-collinear ground state of the dimer and the possibility of the existence of multiple magnetic solutions for the same parameters of the model. The fifth section examines the possibility of controlling the effective magnetic interaction by varying the external magnetic field, which allows for a new interpretation of the STM experiments. The final section briefly summarizes the  results of the study.

\section{Magnetic dimer in NC AA model}
Initially, the AA model was formulated for the description of two interacting magnetic impurities in the non-magnetic metallic matrix~\cite{alexander1964interaction} 
In the mean-field approximation for Coulomb repulsion on-site, this model is a variant of tight-binding theory. It reproduces many of the results of density functional calculations \cite{Oswald_JPF_85} and to some extent takes into account the effects of electron correlation~\cite{Katsnelson_PRB_2000}.
The AA Hamiltonian describes the d-impurities in the sea of quasi-free s(p)-electrons of the conduction band. It can be written as
\begin{equation}\label{1}
\begin{aligned}
H&=\sum_{\bf k,\alpha}\epsilon_k n_{\bf k\alpha}+\sum_{i,\alpha}\epsilon_i^0 n_{i\alpha}+\sum_{{\bf k},i,\alpha}(\upsilon_{i{\bf k}} d_{i\alpha}^{\dagger} c_{{\bf k}\alpha}+\upsilon_{{\bf k} i} c_{{\bf k}\alpha}^{\dagger} d_{i\alpha})\\
&+\sum_{i \neq j,\alpha}\upsilon_{ij} d_{i\alpha}^{\dagger} d_{j\alpha}+\frac{1}{2}\sum_{i,\alpha}U_i n_{i\alpha} n_{i-\alpha}
\end{aligned}
\end{equation}
Here $d_{i\alpha}^{\dagger}$($d_{i\alpha}$) are the  creation (annihilation) operators of d-electrons with spin $\alpha$ ($\alpha=\pm 1$) localized on impurity $i$ ($i=1,2$) with energy $\epsilon_i^0$ while $c_{\bf k\alpha}^{\dagger}$ ($c_{{\bf k}\alpha}$) are the corresponding operators for s(p)-electrons with momentum $\bf k$ and energy $\epsilon_k$ in the conduction band;
$n_{i\alpha}=d_{i\alpha}^{\dagger}d_{i\alpha}$ and $n_{\bf k\alpha}=c_{ \bf k\alpha}^{\dagger}c_{\bf k\alpha}$ are the operators of occupation numbers.
The parameters of s(p) -d hybridization and direct electron hopping between impurities $i$ and $j$ are denoted by $\upsilon_{{\bf k}i}$ and $\upsilon_{ij}$, respectively. The value $U_i$ defines the Coulomb repulsion of electrons localized on impurity $i$.

For the d-subsystem, the presence of the s(p)-conduction band leads to a finite width $\Gamma$ of   d-levels and a shift in their energy, as in the single-impurity Anderson model \cite{Anderson_PR_1961}. 
The value of $\Gamma$ for 3d impurities in a metallic matrix is typically 0.5–1 eV. If this value exceeds the splitting of the d-levels by the crystal field, then the d-states can be considered fivefold degenerate and equally filled.
\begin{equation}
\begin{aligned}
E_i^0=\epsilon_i^0 + \rm{Re} \sum_{\bf k} \frac{\upsilon_{i \bf k}\upsilon_{{\bf k}i}}{\varepsilon-\epsilon_k},  \text{ }
    \Gamma = \rm{Im} \sum_{\bf k} \frac{\upsilon_{i \bf k}\upsilon_{{\bf k}i}}{\varepsilon-\epsilon_k},
    \end{aligned}
\end{equation}

The hopping parameter~($V_{ij})$ is also renormalized due to transitions through the conduction band:

\begin{equation}
\begin{aligned} 
    V_{ij}=\upsilon_{ij}+\sum_{\bf k} \frac{\upsilon_{i \bf k}\upsilon_{{\bf k}j}}{\varepsilon-\epsilon_k}
    \end{aligned}
\end{equation}
To describe the non-collinear structure, we use the mean-field approximation in a local reference frame with the quantization axis along the magnetic moment on each impurity \cite{Hirai_JPSJ_1992,Uzdin_Comp-mat-Sci_1998}: 
$n_{i\alpha}n_{i-\alpha}=<n_{i-\alpha}>n_{i\alpha}+<n_{i\alpha}>n_{i-\alpha}-<n_{i\alpha}><n_{i-\alpha}>$.
After transitioning to a common laboratory reference frame for both impurity atoms with  quantization axis forming angles $\theta_i$ and $\phi_i$ (i=1,2) with the directions of the magnetic moment, we obtain the effective Hamiltonian $H_d^{nc}$ for the d-subsystem.
\begin{equation}\label{3}
H_d^{nc}=\sum_{j,\alpha}E_j^{\alpha}n_{j\alpha}+\sum_{j,l,\alpha,\beta}V_{jl}^{\alpha\beta}d_{j\alpha}^{\dagger}d_{l\beta}-\frac{1}{4}\sum_j U_j(N_j^2-M_j^2),
\end{equation}
where
\begin{equation}\label{4}
\left\{
\begin{aligned}
E_j^{\alpha}&=E_j^0+\frac{U_j}{2}(N_j-\alpha M_j{\rm{cos}}\theta_j)\\
        V_{jl}^{\alpha \beta}&=\frac{U_j M_j}{2}(\delta^{\alpha \beta}-1)\delta_{jl}e^{-\alpha i \phi_j}{\rm{sin}} \theta_j +(1-\delta_{jl})\delta^{\alpha \beta}V_{jl}
\end{aligned}
\right.
\end{equation}
Here, the indices $j$ and $l$ label the impurity atoms ($j,l=1,2$). The Greek indices $\alpha$ and $\beta$ correspond to the spin variables.
The number of d-electrons $N_j$ and the magnitude of the magnetic moments $M_j$ in Eq.~\eqref{3} and Eq.~\eqref{4} can be expressed in terms of the Green's function using the following relations \cite{Bessarab_PRB_2014, bessarab2014navigation}:

\begin{equation}\label{5}
\begin{aligned}
N_j&=\frac{1}{\pi}\int_{-\infty}^{\epsilon _F} \rm{d}\epsilon\ \rm{Im}[G_{jj}^{++}(\epsilon-i\Gamma)+ G_{jj}^{--}(\epsilon-i\Gamma)], \\
M_j&=\frac{1}{\pi}\int_{-\infty}^{\epsilon _F} \rm{d}\epsilon\ \rm{Im}[G_{jj}^{++}(\epsilon-i\Gamma)- G_{jj}^{--}(\epsilon-i\Gamma)]cos\theta_j + \\ 
&+ \frac{1}{\pi}\int_{-\infty}^{\epsilon _F} \rm{d}\epsilon\ \rm{Im}[G_{jj}^{+-}(\epsilon-i\Gamma)e^{i\phi_j}+ G_{jj}^{-+}(\epsilon-i\Gamma)e^{-i\phi_j}]sin\theta_j
\end{aligned}
\end{equation}

\begin{figure*}[htb]
    \centering
    \includegraphics[width=1\linewidth]{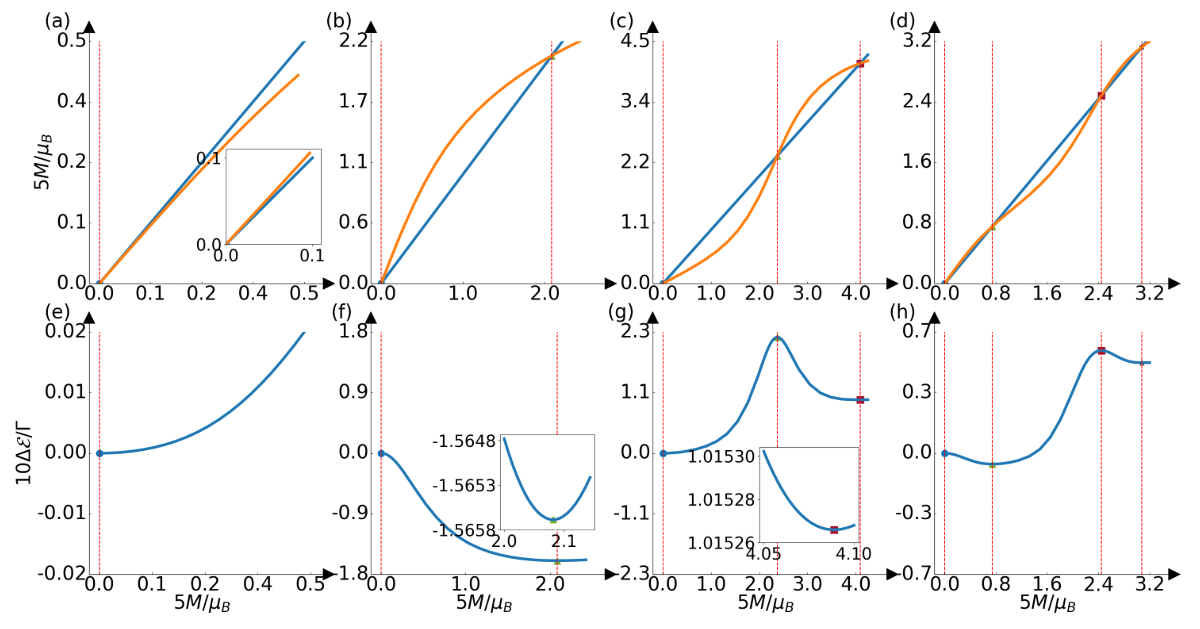}
    \caption{(Color online) 
    Graphical solution of equation (\ref{8}) and the dependence of dimer energy $10\Delta \mathcal{E}/ \Gamma$ on  5$M$. All variants with different numbers of  solutions are shown}

\label{fig1}
\end{figure*}

If we choose the quantization axis in the laboratory system along the magnetic moment of the first atom, then the following expression can be obtained for the Green's functions.
\begin{equation}\label{6}
\begin{aligned}
  G_{11}^{\alpha\alpha}=\frac{1}{\displaystyle \epsilon-E_1^{\alpha}- \frac{V_{12}V_{21}}{\displaystyle \epsilon-E_2^{\alpha}-\frac{V_{22}^{\alpha - \alpha}V_{22}^{-\alpha\alpha}}{\displaystyle\epsilon-E_2^{-\alpha}-\frac{V_{12}V_{21}}{\displaystyle\epsilon -E_1^{-\alpha}}}}} \\
 %   G_{11}^{\alpha-\alpha}=  G_{11}^{-\alpha\alpha}=0
  \end{aligned}
\end{equation}
The density of states (DOS) with each spin projection is now a superposition of at most four Lorentz contours of width $\Gamma$. We will consider a dimer from 2 identical atoms: $E_1^0=E_2^0\equiv E^0, U_1=U_2\equiv U , V_{12}=V_{21}\equiv V$. The number of d-electrons and magnetic moment on both atoms are also assumed to be the same $N_1=N_2=N, M_1=M_2=M$.
All energy parameters below will be measured in units of $\Gamma$. In the collinear case, the DOS contains only two contours. For parallel moment ordering, their positions are separated by twice the hopping parameter $V/\Gamma$, and their amplitudes are the same. The contours for different spin projections are shifted by $UM/2\Gamma$. For antiparallel moment ordering, the positions of contours with different spins coincide, but their amplitudes are different.

In the general case equations for the self-consistent determination of the number of d-electrons $N_1$ and the magnitude of the magnetic moment $M_1$ %localized on the atoms of a magnetic dimer (\ref{5}) 
can be written as
\begin{equation}\label{7}
N={\sum_{\alpha}}\:{\sum_{\mu=1}^4} \frac{p_\mu^\alpha}{\pi}
\rm{arccot\frac{q_\mu^\alpha-\epsilon_F}{\Gamma}}
\end{equation}
\begin{equation}\label{8}
M={\sum_{\alpha}}\:{\sum_{\mu=1}^4} \frac{\alpha p_\mu^\alpha}{\pi}
\rm{arccot\frac{q_\mu^\alpha-\epsilon_F}{\Gamma}}
\end{equation}

In this expression $q_\mu^\alpha$ are the roots of the denominator of Green's function Eq.~\eqref{6}, and $p_\mu^\alpha$ are the coefficients of its expansion into simple fractions. These quantities depend on $M$ and $N$, as well as on the model parameters $x=\frac{E^0-\epsilon_F}{\Gamma}$, $y=\frac{U}{\Gamma}$, $v=\frac{V}{\Gamma}$. 
%Therefore, we have already obtained the equations of self-consistency.
Therefore, (\ref{7}) and (\ref{8}) are {%indeed} 
the equations of self-consistency.

%\emph{Non-collinear solutions}
If the magnetic moment $M$ and the number of particles $N$ are given, the energy of the system can be found even if $M$ and $N$ differ from the equilibrium values. 
To do this, it is sufficient to integrate the DOS to the Fermi level and subtract the last term in Eq.~\eqref{3}, obtained in the mean field approximation.
We define 
\begin{equation}\label{9}
\begin{aligned}
    \mathcal{E}(M,N)=\frac{1}{\pi}\sum_{\alpha}\sum_{\mu=1}^{4}p_\mu^\alpha\Big[q_\mu^\alpha\rm{arccot\frac{q_\mu^\alpha-\epsilon_F}{\Gamma}}+ \\ +\frac{\Gamma}{2}\rm{ln}\Big( \frac{(q_\mu^\alpha-\epsilon_F)^2}{\Gamma^2}+1\Big) \Big]-\frac{U}{4}(N^2-M^2)
\end{aligned}
\end{equation}
where all notations are the same as in right hand Eqs. (\ref{7}),(\ref{8}). Then $ \Delta\mathcal{E}(M,N)= \mathcal{E}(M,N)- \mathcal{E}(0,N)$ represents the energy of a state with magnetic moment $M$, measured from the nonmagnetic state with the same d-electron number $N$, because contributions from the lower limit of DOS integration in the magnetic and nonmagnetic states will cancel out.  Accounting for the fivefold degeneracy of the d-levels leads to a fivefold increase in the total number of particles $N$ (\ref{7}) and the magnetic moment $M$ (\ref{8}) on each atom, as well as a tenfold increase in the total dimer energy $\Delta \mathcal{E}$.

When the angle $\theta$ between the magnetic moments changes, the energy surface defined by expression Eq.~\eqref{9} is also transformed. If we assume that the position of the d-level relative to the $\epsilon _F$ defined by the parameter $x$ remains unchanged, then the number of d-electrons will adjust and the equilibrium state will correspond to the minimum of the grand canonical potential $\Delta \Omega=\Delta(\mathcal{E}-\epsilon_F N)$. In this case, self-consistent solutions will correspond to extrema (minima or maxima) on the $\Delta \Omega (N,M)$-surface. Indeed, the conditions $(\frac{d\Delta \Omega}{dN})_M=0$ and $(\frac{d \Delta \Omega}{dM})_N=0$ coincide with the self-consistency equations Eq.~\eqref{7} and Eq.~\eqref{8}, respectively.

If the number of particles is fixed, $N=N^*$, then changing the angle will adjust the parameter $x$, and the equilibrium states with a given angle $\theta$ correspond to the extrema of the canonical potential.

In this case, we can introduce the function $\Delta \Omega(M,x) = \Omega(M,x(N^*,M))-\Omega(0,x(N^*,0))$
where $x(N^*,M)$ determines the position of d-level that will give self-consistent values of $N^*$ (Eq. (\ref{7})) for a given magnetic moment $M$. If we relate the change in parameter $x$ to the shift of the Fermi energy with respect to $E^0$, we obtain 
$$\Delta \mathcal{E} (M,x)= \Delta \Omega(M,x)-\Gamma N^*[x(N^*,M)-x(N^*,0)]$$
Then condition {$(\frac{d\Delta \mathcal E}{dM})_x=0$ yields the self-consistency condition (9), and  $(\frac{d\Delta \mathcal E}{dx})_M=0$ will provide the number of d-electrons equal to $N^*$.

Thus, the d-subsystem within the NC AA model can be considered both in the canonical ensemble, where the energy is considered as a functional of the magnetic moment and the position of the d-level relative to the Fermi level, and in the grand canonical ensemble, where the grand canonical potential is considered as a functional of the magnetic moment and the number of d-particles. Both approaches can be used to describe real systems.

For example, when considering magnetic dimers of different 3d metal atoms, the canonical ensemble proves more convenient. The number of d-electrons in the system does not necessarily have to be an integer, but should roughly be  equal to the number of electrons in an isolated atom. This ensures the condition of electroneutrality. The value of the d-level $x$ will be determined by the required number of d-electrons, taking into account that these electrons hybridize with s(p) electrons of the matrix.

 When studying the effect of an external electric field (for example, created by a scanning tunneling microscopy (STM) tip) on the state of a specific magnetic dimer, it is more convenient to consider it in a grand canonical ensemble. The electric field causes a change in the position of the d-level relative to the Fermi level, determined by the concentration of free electrons in the matrix. This, in turn, changes the number of localized d-electrons in the dimer. However, these effects are interrelated, and the two states can be considered as the states of two different dimers with different numbers of electrons before and after the field is applied and, accordingly, with different positions of the d-levels relative to the Fermi level.

\section{Self-consistent solutions in NC AA model}

Let us now consider possible solutions of Eq.~\eqref{7} and \eqref{8} in the canonical ensemble for various number of d-electrons $N$. The corresponding data for different  positions of the d-level $x$ in the grand canonical ensemble are given in the appendix. To find self-consistent solutions, note that the right-hand side of Eq.~\eqref{7}  is a monotonically decreasing function of $N$, while the left-hand side is a monotonically increasing function.
Therefore, this equation has a unique solution. Substituting this solution into  Eq.~\eqref{8}, we can graphically find possible self-consistent values of $M$ as shown on the upper panel of Fig.~\ref{fig1}. All possible variants of solutions are shown in Fig.~\ref{fig1} along with the dependence $\Delta \mathcal{E}(M)$.  The dependencies are plotted for the already self-consistent values of $x$ for the correspondent $M$. 

 In accordance with the above, the extrema of $\Delta \mathcal{E}(M)$ in Fig.~\ref{fig1} correspond to self-consistent solutions. Note that there always exists a non-magnetic solution. 
The case when it is the only solution is shown in Fig.~\ref{fig1}~(a, e). In this case, adopting the parameters ${5}N=8.5$, $\theta=\pi/4$, and $v=2$, we find that the self-consistent solution occurs at $M=0$, with a corresponding value of $x = -14.13$.
    \begin{figure}[tb]
    \centering
    \includegraphics[width=1\linewidth]{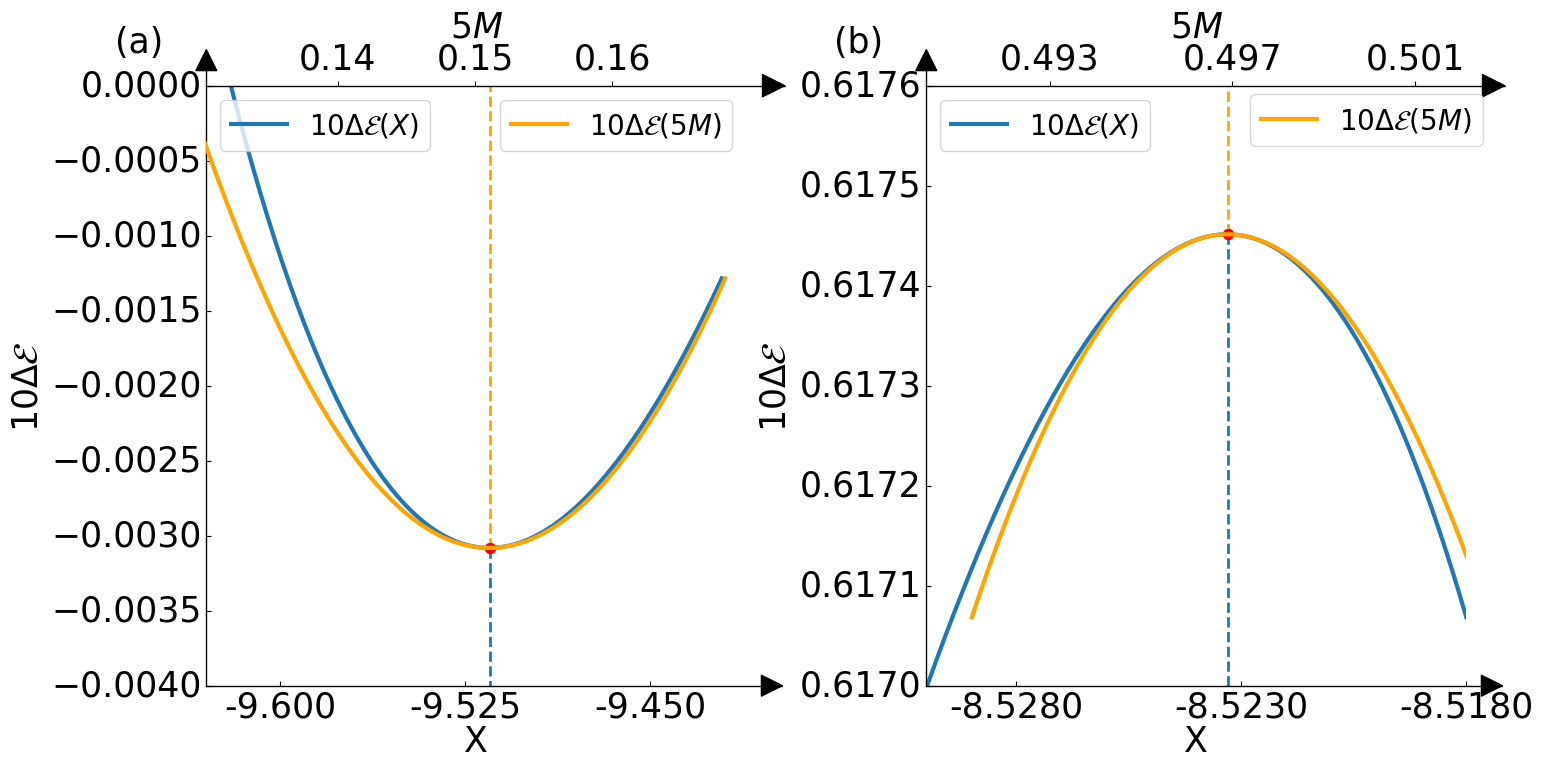}
    \caption{(Color online) Dimer energy $10\Delta \mathcal{E}/\Gamma$ as a function of $x$ and $M$ near the self-consistent solution for the states shown in  Fig.~\ref{fig1}(d),(h). For $\Delta \mathcal{E}(M)$, the $x$ is determined self-consistently using eq.~\eqref{7} to ensure specified number of d-electrons $5N=6$.  When calculating dependence $\Delta \mathcal{E}(x)$, $M$ satisfy to eq.~\eqref{7}; (a) represents the stable state under electroneutrality, whereas (b) corresponds to the unstable state.
}
\label{prove}
\end{figure}

\begin{figure*}[!tb]
    \centering
    \includegraphics[width=0.7\linewidth]{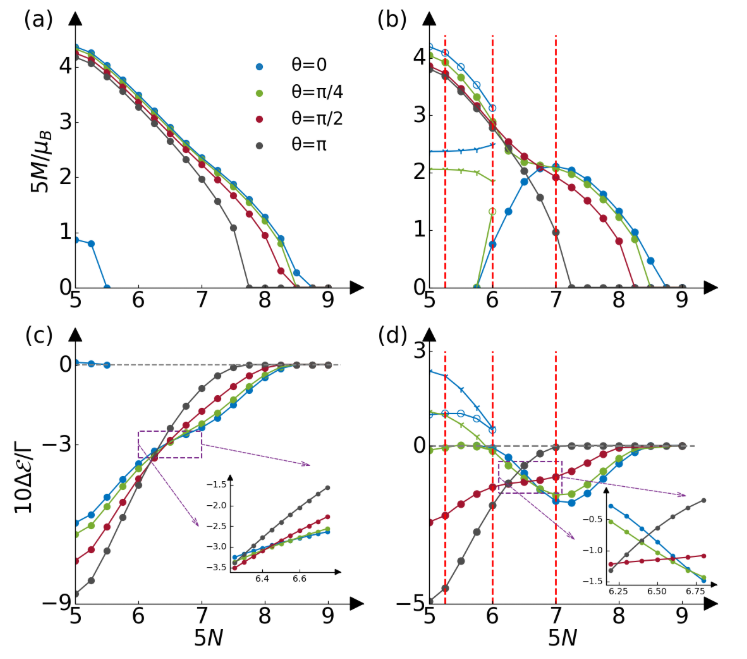}
    \caption{(Color online) Self-consistent magnetic moment per atom 5$M$ and  energy of dimer $10\Delta \mathcal{E} /\Gamma$ as functions  of $d$-electron numbers 5$N$.
    Parameter values: $ y=13$ (a,c) $v=2$, (b,d) $v=3$.
    Insets shows the region of non-collinear ground state}
    \label{fig2}
\end{figure*}
In Fig.~\ref{fig1} (b, e) for $5N=7$, $\theta=\pi/4$, and $v=3$, one non-magnetic solution $M=0$ ($x=-11.95$) and one magnetic solution $\textcolor{blue}{5}M=2.10$ ($x=-11.77$) are presented. The non-magnetic solution corresponds to the maximum of $\Delta \mathcal{E}(M)$ and is unstable. The magnetic solution represents the ground state for a given $N$ and self-consistently determined $x$. 

Figs.~\ref{fig1} (c, f) show the case where 3 self-consistent solutions are possible. The non-magnetic solution is the ground state of dimer. The magnetic solution with a small moment (about 1 $\mu_B$) is unstable, and the second magnetic solution with a moment greater than 4 $\mu_B$ is locally stable. For the case of $\textcolor{blue}{5}N=5.25$, $\theta_2=0$, and $v=3$, self-consistent solutions appear at the points $(M, x) = (0, -7.57)$, $(2.35, -6.98)$ and $(\textcolor{blue}{5}M, x) = (4.05, -7.67)$.

Fig.~\ref{fig1} (d, h) show the case where there are four self-consistent solutions of Eq.~\eqref{8}. The non-magnetic solution and the magnetic state with momentum 2.4 $\mu_B$ correspond to the maximum $\Delta \mathcal{E}(M,N)$ and are unstable. The other two magnetic solutions are locally stable. In the ground state the moment is lower. These four distinct self-consistent solutions were obtained for  $\textcolor{blue}{5}N=6.0$, $\theta_2=0$, and $v=3$. They can be expressed as pairs of $(\textcolor{blue}{5}M, x)$: $(0, -9.87)$, $(0.75, -9.52)$, $(2.5, -8.52)$, and $(3.15, -8.93)$.

  The stability of self-consistent solutions is illustrated in the bottom panel of Fig.~\ref{fig1}, which shows the dependencies $10\Delta \mathcal{E}/\Gamma (M)$, where the parameter $x$ is chosen to ensure a given self-consistent numbers of d-electrons $N$.
 If $N$ differs from the value determined by the nuclear charge, the impurity will be charged, increasing the energy of the system, although this contribution is not explicitly included in the Hamiltonian of the d-subsystem \eqref{1}. Assuming that the charge (number of particles) relaxation time is significantly shorter than the magnetic moment relaxation time, one can conclude that the state will be stable if it corresponds to a minimum in $M$. 

 The same is true for the maximum of $\Delta \mathcal{E}$. This is shown in Fig.~\ref{prove}, where the energy is plotted as a function of $x$ and $M$ in the neighborhood of the two self-consistent solutions shown in Fig.~\ref{fig1}(d). The dependence $10\Delta \mathcal{E} (M)/\Gamma$ is obtained for $x$ and $M$ satisfying \eqref{7}. 
 
The fact that an extremum point represents a minimum or maximum under certain conditions (in our case, self-consistency conditions \eqref{7}) does not guarantee that the corresponding points are actually local minima or maxima on the surface  $\Delta \mathcal{E}(M,N)$.
To determine the type of extremum, it is necessary to determine the eigenvalues of the Hessian matrix at the corresponding points. The Hessian matrix can be expressed in terms of the density of states at the Fermi level $\rho^\alpha \equiv \rho^\alpha (\epsilon_F)$ if the quantization axis is chosen along the magnetic moment of the corresponding atom. The equation for the eigenvalues of the Hessian matrix is:
\begin{equation}\label{gessian roots}
\begin{aligned}
    \lambda_{1,2}=\frac{U}{2}\Big[ -\frac{y}{2}(\rho^\alpha+\rho^{-\alpha})
    \pm \sqrt{(\frac{y}{2})^2 (\rho^\alpha-\rho^{-\alpha})^2+1} \Big]
\end{aligned}
\end{equation}
One of the eigenvalues of the Hessian matrix is always negative, while the other can be positive or negative. Thus, the extrema represent saddle points or maxima. The minima shown in the figure correspond to saddle points, and the maxima correspond to maxima. The criterion for an eigenvalue to cross zero corresponds to the Stoner criterion for the transition from a non-magnetic to a magnetic state in the systems with itinerant electrons. A saddle point can become a minimum if the electrostatic energy responsible for electroneutrality is added.

Fig.~\ref{fig2} shows the dependencies of the self-consistent magnetic moment $5M$ and dimer energy $10\Delta\mathcal E/\Gamma$ on the number of d-electrons per atom $5N$ for different angles $\theta$. Figs.~\ref{fig2} (a, c) and (b, d) were obtained for $v$ = 2 and $v$ = 3, respectively. The dependencies are symmetric with respect to the point $N$=1:
$M(10-5N)=M(5N)$, $\Delta \mathcal{E} (10-5N)=\Delta \mathcal{E} (5N)$. Therefore, only the range of values for $5N>5$ is shown.
The DOS on the first atom for symmetric states satisfies the relation $\rho_1(\epsilon-\epsilon_F) = \rho_1(\epsilon_F-\epsilon)$. Only region $5N>5$ is presented in Fig.~2.

For both $v$=2 and $v$=3, there is a domain where the ground state is non-collinear. For $N>5$ to the left of this region, the ground state is AF with the maximum AF exchange at $N$=5. To the right, the ground state is FM. This is consistent with the dependence of the exchange parameter on the Fermi level reported in \cite{Liechtenstein_JPF_1984}.

\vspace{0.5cm}

\section{Non-collinear ground state and plural solutions}

For certain values of the parameters in the tight-binding model, a canted ground state of the magnetic dimer was obtained in \cite{costa_PRL_2005} for closely located magnetic impurities. With increasing distance  the interaction became collinear FM or AF. In our case, the noncollinearity 
behavior does not change qualitatively with variation of the hopping parameter $v$, although the strength of the effective interaction and the values of moments decrease with increasing $v$.

In the domain of  noncollinear ground state, the effective exchange stiffness is small, and spontaneous emergence of noncollinear localized magnetic states without DMI and spin-orbit interaction is possible here. Realignment of such structures by a small external action may be of interest for the development of new magnetic memory circuits and other spintronics applications.

\begin{figure}[!tb]
    \centering
    \includegraphics[width=0.8\linewidth]{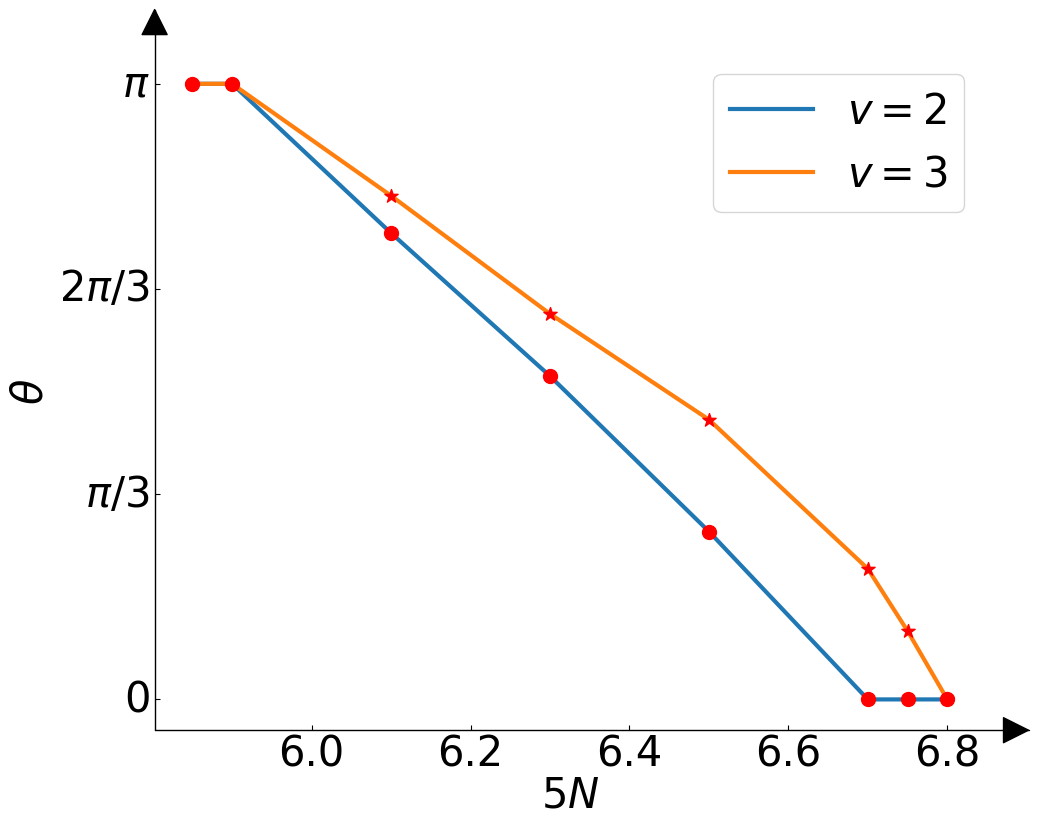}
    \caption{(Color online) Angle $\theta$ between the magnetic moments corresponding to the minimum of the dimer energy as a function of the d-electrons number $5N$ for two values of the hopping parameter, $v=2$ and $v=3$.}

    \label{figtheta-N}
\end{figure}

To determine the magnetic ground state, we minimize the total energy with respect to the angle $\theta$ between the magnetic moments for
each value of the electron number $N$. The resulting dependence of the optimal angle $\theta$ on $5N$ is shown in Fig.~\ref{figtheta-N} for two representative values of the hopping parameter $v$. When the number of electrons is close to the half-filled d-band ($5<5N \lesssim 6$ in Fig.~\ref{figtheta-N}), the energy minimum is located at $\theta = \pi$, which corresponds to the antiferromagnetic state.
 As $N$ increases, the minimum shifts continuously
toward smaller angles, indicating the formation of non-collinear
magnetic states. Finally, for sufficiently large electron numbers
($5N \gtrsim 6.7$), the minimum occurs at $\theta = 0$, corresponding
to a ferromagnetic alignment. This continuous evolution of $\theta$ demonstrates that the magnetic
ground state is obtained by a full minimization over the angle and
confirms the existence of a non-collinear regime between the
antiferromagnetic and ferromagnetic limits.

\begin{figure}[b]
    \centering
    \includegraphics[width=1.0\linewidth]{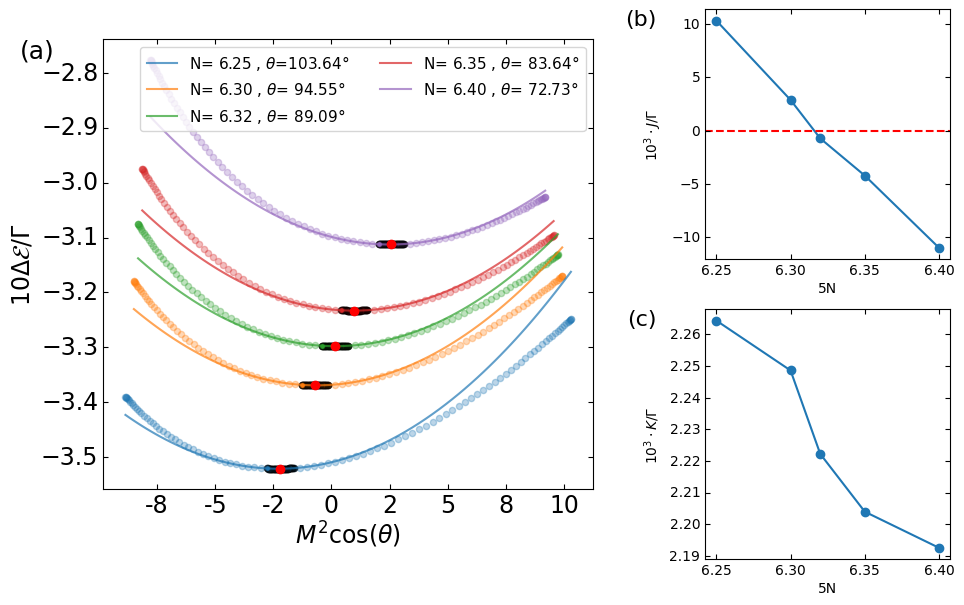}
    \caption{(Color online) Left (a): Total energy as a function of $M^2\cos(\theta)$ for several values of the electron number $N$ close to the crossover region. The symbols denote the results obtained from the AA model, while the solid lines represent fits using the effective model including bilinear ($J$) and biquadratic ($K$) exchange interactions. The red dots indicate the positions of the energy minima. Right: Extracted values of the effective exchange parameters $J/\Gamma$ (b) and $K/\Gamma$ (c) as a function of $5N$. }
    \label{figfitJ}
\end{figure}

The dependence of the energy on the angle $\theta$ between the dimer's magnetic moments at a given d-electron number $N$ allows us to determine the effective parameters of the Heisenberg model, which give similar behavior at small  deviations of magnetic moment from the equilibrium state. In the region of the noncollinear ground state, the bilinear contribution alone is clearly insufficient, and the biquadratic interaction must be included. Let us write the dependence of energy on the angle and the magnitude of the magnetic moment in the form
\begin{equation} \label{fit}
E(\theta) = E_0 + J M^2\cos\theta + K(M^2\cos\theta)^2 .
\end{equation}
Here $E_0$ is the constant reference energy, $J$ and $K$ are the bilinear and biquadratic exchange parameters, respectively. Within this framework, the magnetic interaction between the localized moments is fully determined by the dependence of the total energy on
the variable $M^2\cos(\theta)$.

To further analyze the effective magnetic interactions, we focus on the parameter regime near the crossover between FM and AF magnetic configurations. In Fig.~\ref{figfitJ}, we plot the total energy as a function of
$M^2\cos(\theta)$ for several values of $N$ close to this region and
fit the results using (\ref{fit})

The fitted parameters $J$ and $K$ are shown in the right panels of
Fig.~\ref{figfitJ}. As the electron number $N$ varies, the bilinear exchange
interaction $J$ changes sign and passes through zero at a critical
value of $N$, while the biquadratic term $K$ remains finite and varies
smoothly.
This behavior indicates that in the vicinity of $J \approx 0$, the
magnetic interaction is no longer dominated by the bilinear exchange
term, but instead by the biquadratic contribution. As a result, the
energy landscape becomes relatively flat with respect to
$M^2\cos(\theta)$ near its minimum, leading to an enhanced sensitivity
to angular fluctuations and favoring non-collinear magnetic
configurations. 

It should be noted, however, that adding magnetic moments to the system (for example, going from a dimer to a trimer) generally changes the values of these pairwise interaction parameters, i.e., the magnetic interaction ceases to be pairwise. However, this is beyond the scope of current study.

Let us now consider the possibility of the existence of several locally stable solutions with different values of the magnetic moments. This qualitatively distinguishes the system under consideration from Heisenberg-type models. For $v$=2, several stable solutions are not realized. Only for $\theta=0$ does a second self-consistent solution exist, corresponding to the energy maximum, as shown  in Fig.~\ref{fig1}  (c). However, the non-magnetic state here is unstable with respect to the formation of AF order in the absence of magnetic fields.

At $v$ = 3, there are multiple solutions for $N$ located between regions with noncollinear ordering. The presence of several solutions occurs with parallel ordering of moments or with small angles $\theta$ between them.  In this case, states with a large angle $\theta$ or antiparallel ordering are energetically more favorable. Nevertheless, when a magnetic field is turned on that stabilizes states with small $\theta$, such states can be observed, and transitions between them are possible when the field changes. Note that states with an intermediate magnetic moment, which are practically independent of the number of particles, correspond to a local energy maximum. All the possibilities shown in Fig.~\ref{fig1} are realized in Fig.~\ref{fig2}  and are marked there with dotted vertical red lines. On the basis of this finding, we conclude that the number of d-particles has a pronounced effect on the exchange interactions, as evidenced by the threefold change in the energy difference.

\vspace{1cm}

\section{Discussion}

The possibility of multiple solutions in transition metal nanoclusters on a Cu surface was demonstrated using DFT calculations. However, this was obtained only for clusters larger than a dimer  with different electronic states on different atoms \cite{stepanyuk_jmmm_1997}. Experiments and calculations on a  Co atom on a black phosphorus surface  demonstrated the possibility of multiple magnetic states even for a single atom, with different electron configurations of its various orbitals \cite{kiraly_Nat-comm_2018,badrtdinov_2D-mater_2020}. In all cases, the authors claim the possibility of controlling the different magnetic states using STM. Non-collinear states and multiple magnetic solutions  which we found are realized already in 3d-dimer with degenerate d-states.

The dependence of the exchange interaction on the number of particles localized in the dimer atoms allows for a new interpretation of the magnetization reversal mechanisms for magnetic structures observed in STM experiments. They show that to stimulate magnetization reversal, the current $I$ from the STM tip (and the voltage $U$ between the tip and the surface) must be sufficiently large \cite{krause_Science_2007,loth_Science_2012,khajetoorians_Science_2013,romming_Science_2013}. The effect is not limited to the release of Joule heat and magnetic field produced by current \cite{krause_Science_2007}, since at constant power, switching rates depend critically on $U$ and significantly less on $I$ \cite{romming_Science_2013}. The underlying mechanism is believed to be related to the moments of forces acting on the magnetic structure during current flow. This mechanism explains the different magnetization switching rates for current flowing from the tip to the surface and in the opposite direction, but without a quantitative description of the experiment~\cite{krause_Science_2007,romming_Science_2013}.

The results obtained above suggest another possible mechanism for stimulating the magnetic transition associated with the annihilation of magnetic skyrmions. As the STM tip approaches the surface, the corresponding electric field affects the position of Fe d-levels atoms relative to the Fermi level and consequently the number of d-electrons $N$. With increasing $N$, the energy of the magnetic state decreases relative to the non-magnetic one, as shown in Fig.~\ref{fig2}, making the structure less stable to thermal fluctuations. When a voltage is applied in the opposite direction, the number of d-electrons $N$ decreases, and the energy of the ferromagnetic state becomes lower relative to the non-magnetic state. However, for Fe, which has approximately 7 d-electrons (i.e., $5N \approx 7$), the bilinear exchange interaction decreases, and the system becomes unstable to changes in the direction of the moments. These effects are not symmetric with respect to the direction of the electric field, but in both cases they will lead to a decrease in the lifetime of the magnetic states, as observed experimentally. It should be noted that the main effect is related to the magnitude of the voltage (electric field strength) and can be observed even without a tunneling current. In principle, this can be verified experimentally.
As mentioned above, the energy scale in our theory is determined by the parameter $\Gamma$, which for 3d elements is approximately 0.5–1 eV. In the STM experiment, the potential difference between the positive and negative potentials of the tip (600 meV and -600 meV) exactly corresponds to this energy scale, and from this point of view, the theory agrees with the experimental data.

More generally, the proposed scenario allows changing the magnetic exchange interactions and the local magnetic structure by applying an electric field that changes the energy of the d-level relative to the Fermi level determined by the s(p) conduction electrons.
 Control of magnetism by electric fields has attracted recently considerable interest  \cite{Negulyaev_PRL_2011,fert_Rev-Mod-Ph_2024}. It can be performed  by modifying interfacial charge, orbital hybridization, and exchange interactions; in ferroelectric heterostructures such as BiFeO$_3$-based systems, polarization switching has been shown to reversibly tune exchange bias and reorient interfacial magnetic order~\cite{wu2010reversible,heron2014electric}.  Suggested mechanism demonstrates another mechanism of manipulation by magnetic interactions using electric field .

Finally, we note that the model under consideration does not take into account a number of interactions that may be significant under certain conditions for description of 3d structures. For example, the mean-field approximation does not account for the electron correlations responsible for the Kondo effect at low temperature. The model under consideration neglects the splitting of the d-levels by the crystal field, leading to a more complex shape of the density of states. The approximation in which the fivefold degeneracy of the d-levels is reduced to a renormalization of the parameters of the Coulomb and exchange interactions  is justified when the splitting of the levels is small compared to their width due to the s-d hybridization of $\Gamma$. This approximation was used and discussed in the original works of Anderson \cite{Anderson_PR_1961,alexander1964interaction}, where it was stated that, despite their simplicity, the models contain significant physics. Moreover, in a number of subsequent works in which calculations were performed by the DFT method, taking into account a large set of interactions, the AA model was used to interpret the obtained results and explain the physics of magnetic states studied in ab initio approach \cite{Oswald_JPF_85, Lounis_PRB_2005}. 
We also note that the noncollinear AA model has been used to explain spin-resolved M\"ossbauer spectroscopy  results 
\cite{Uzdin2012}, magnetization reversal of Fe islands on a W surface\cite{Bessarab_PRB_2014}, and magnetization reversal of clusters at the tip of a tunneling microscope \cite{ivanov2017magnetic}. Thus, despite the relative simplicity of the model and the fact that it contains a very small number of phenomenological parameters, such models can be quite useful for quantitative analysis of experimental data and for predicting new effects.

\section{conclusion}

In conclusion, we demonstrated that within the non-collinear AA model for itinerant electrons, the interaction of localized states strongly depends on the occupation numbers of d-states. Self-consistent solutions correspond to extreme points on the energy surface as functions of d-electron number and the magnetic moment. Regions of non-collinear magnetic ordering in a magnetic dimer on a nonmagnetic substrate and regions of the existence of several solutions with different magnetic moments were found. Our work highlights two key findings. First, we demonstrate that the d-electron number significantly controls the exchange interaction within the system, manifested as changing the angle between magnetic moments in the ground state. Second, we uncover a new mechanism for magnetization reversal in nanostructures driven by an external electric field coming from a tip in STM experiments.

\vspace{1cm}
\section*{Acknowledgment}
This work is supported by the Russian Science Foundation (Grant no. 23-72-10028), the Shenzhen Fundamental Research Program (No.~JCYJ20250604145655074), the National Natural Science Foundation of China~(Grant No. 12204130), Shenzhen Key Laboratory of Advanced Functional Carbon Materials Research and Comprehensive Application (Grant No. ZDSYS20220527171407017).

\appendix
\section{Results in grand canonical ensemble}
\begin{figure*}[htb]
    \centering
    \includegraphics[width=0.87\linewidth]{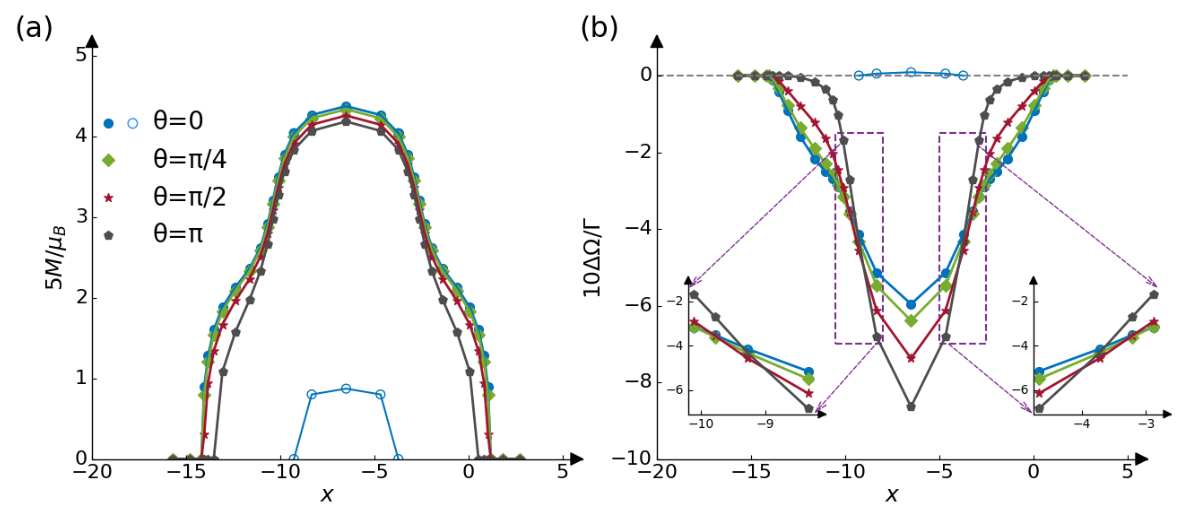}
    \caption{(Color Online) The magnetic moment of each atom and the grand canonical potential of the magnetic dimer as a function of x. Values of parameters are $v=2,y=13$. Individual colors indicate the angles between the magnetic moments. Filled and empty symbols correspond to the ground and unstable states, respectively. The inset shows the non-collinear ground state regions at a larger scale.}
    \label{suppfig1}
\end{figure*}

\begin{figure*}[htb]
    \centering
    \includegraphics[width=0.87\linewidth]{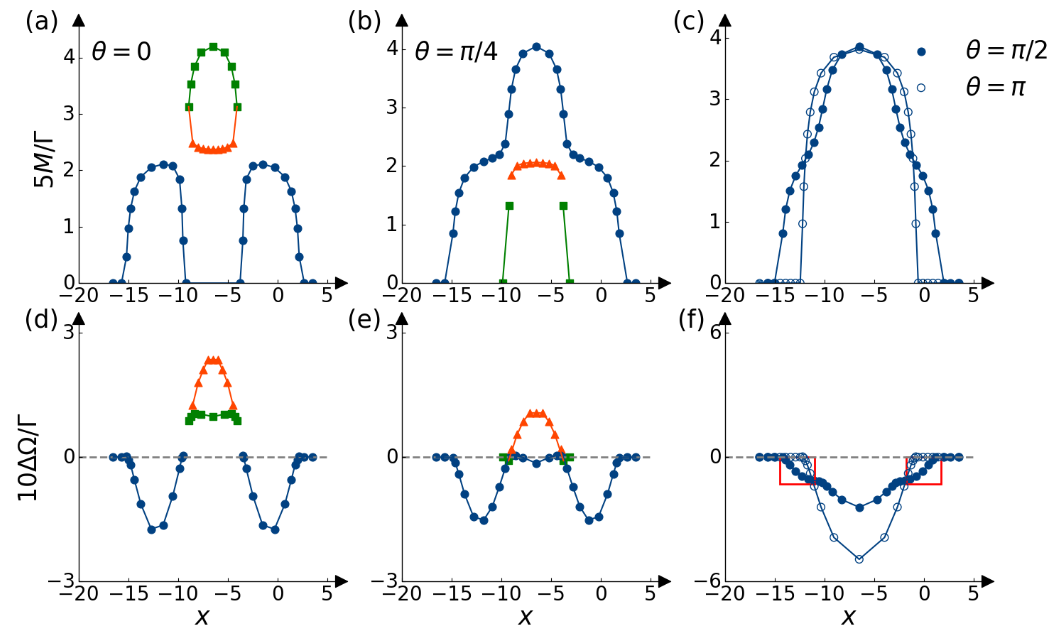}
    \caption{(Color online) Dependence of the magnetic moment per atom and the grand canonical potential of a magnetic dimer on the parameter $x$ for different angles $\theta$: (a),(d) -- $\theta=0$; (b),(e) -- $\theta=\pi/4$, (c),(f) -- $\theta=\pi/2$ and $\theta=\pi$.  Parameter values: $v=3, y=13$.}
    \label{suppfig2}
\end{figure*}

The main text presents the dependencies of the magnetic moment $M$ and the energy $\Delta \mathcal E$ (the canonical potential) of the magnetic dimer relative to its non-magnetic state with the same number of d-electrons. In this case, the self-consistent solutions corresponded to the extrema of $\Delta \mathcal E$, considered as a function of the magnetic moment $M$ and the parameter $x$, which determines the position of the d-level relative to the Fermi level.

The dependencies of the magnetic moment $M$ and the grand canonical potential $\Delta \Omega $ on $x$ are given here. The self-consistent solutions correspond to the extrema of $\Delta \Omega (M, N)$ with respect to $M$ and $N$. Fig.~\ref{suppfig1} shows the dependencies of $M(x)$ and $\Delta \Omega (x)$ for $v = $ 2 and $y =$ 13. The colors display different angles $\theta$ between the magnetic moments in the dimer.

As can be seen, the magnitude of the moments depends weakly on the angle $\theta$. 
All dependencies are symmetric about the point $x=-6.5$, which corresponds to the half-filled band.
There are regions of non-collinear ordering between the states corresponding to ferromagnetic and antiferromagnetic states; these regions are shown in the insets on a larger scale.

Fig.~\ref{suppfig2} shows the evolution of the $x$-dependence of the magnetic moment of the dimer atoms and the grand canonical potential of the dimer with a change in the angle $\theta$ between the magnetic moments. At small $\theta$ there are several self-consistent solutions for the same value of $x$. Magnetic solutions can correspond to the ground, metastable, and unstable states. The first two correspond to local minima of $\Delta \Omega$, as a functional of $M$ and $x$. They are shown in the figure in blue and green. In the latter case, a local maximum is observed, and the corresponding curve is colored orange. 

For $\theta>\pi/2$, only one magnetic solution is possible. Near the half-filled band, the ground state corresponds to an antiferromagnetic ordering of the moments. However, as can be seen in the fig.~\ref{suppfig2}(f), there is a region with red squares where the ground state is non-collinear.

\vspace{15cm}

\newpage
\bibliographystyle{apsrev4-2.bst}
\bibliography{reference.bib}
\end{document}